\documentclass[aps,prc,twocolumn,showpacs]{revtex4}
\usepackage{epsfig}
\newcommand{\ber}{\begin{eqnarray}} 
\newcommand{\eer}{\end{eqnarray}}
\newcommand{\lqcd}{\Lambda_{\rm QCD}}
\newcommand{\mva}{{\overline {v_{2,\alpha}}}}
\newcommand{\mvb}{{\overline {v_{2,\beta}}}}
\newcommand{\mvd}{{\overline {v_{2,\gamma}}}}
\newcommand{\mza}{{\bar z_\alpha}}
\newcommand{\mzi}{{\bar z_i}}
\newcommand{\pt}{p_\perp } 
\newcommand{\pta}{p_{\perp \alpha}} 
\newcommand{\ptb}{p_{\perp \beta}} 
\newcommand{\pti}{p_{\perp i}} 
\newcommand{\vnt}{{\vec n_\perp }}
\newcommand{\vp}{{\vec p}}
\newcommand{\vpa}{{\vec {p_\alpha}}}
\newcommand{\vpb}{{\vec {p_\beta}}}
\newcommand{\vpt}{\vec {p_\perp}}
\newcommand{\vq}{{\vec q}}
\newcommand{\vqa}{{{\vec {q_\alpha}}^{\,\prime}}}
\newcommand{\vqb}{{{\vec {q_\beta}}^{\,\prime}}}
\newcommand{\vqi}{{{\vec {q_i}}^{\,\prime}}}
\newcommand{\vqrest}{{{\vec q}^{\;\prime}}}

\begin{document}
\title{Quark coalescence and elliptic flow of charm hadrons}
\author{Zi-wei Lin and D\'enes Moln\'ar}
\affiliation{Physics Department, The Ohio State University, 
Columbus OH 43210}
\date{\today}
\begin{abstract}
Elliptic flow of charm hadrons is investigated based on the quark coalescence 
model. Due to the large difference between the charm quark and 
light quark masses, hadrons containing both light and charm quarks 
show a qualitatively different $v_2(p_\perp)$ 
from hadrons containing only light quarks.
Simple relations are proposed to infer quark 
elliptic flow from those of hadrons. 
The effects of the finite momentum spread of hadron wavefunctions 
are also studied, and are found to be small for charm hadrons. 
\end{abstract}
\maketitle

\section{Introduction and Conclusions}
During the early stage of ultra-relativistic heavy ion collisions 
at the Relativistic Heavy Ion 
Collider (RHIC) and the future Large Hadron Collider (LHC), 
a deconfined phase, often called the quark-gluon plasma, 
is expected to be formed. 
As the dense matter expands, 
the parton degrees of freedom in the deconfined phase convert to hadrons.
Quark coalescence is a simple model of hadronization,  
where it is assumed that gluons, which are abundant at high temperatures, 
either convert to quark-antiquark pairs or serve to ``dress'' quarks 
into constituent quarks near hadronization. 
The effective degrees of freedom
are constituent quarks and anti-quarks, and mesons 
form from a quark-antiquark pair, while baryons from three quarks  
according to their valence quark compositions. 
Quark coalescence has been applied to heavy ion collisions 
in the algebraic coalescence rehadronization (ALCOR) and 
microscopic coalescence rehadronization (MICOR) models \cite{alcormicor} 
to describe hadron abundances and in a multiphase transport model (AMPT) 
with string melting \cite{Lin:2001zk} to describe the elliptic flow at RHIC. 
Recently it has been used to address the elliptic flow of hadrons 
of different flavors \cite{Lin:2002rw,Voloshin:2002wa,Molnar:2003ff}, 
and the large $p/\pi$ ratio 
\cite{Voloshin:2002wa,Hwa2,Fries:2003vb,Greco:2003xt} observed at RHIC. 

Elliptic flow, $v_2 \equiv\langle \cos(2\phi)\rangle$, 
the second Fourier moment of the azimuthal momentum distribution, 
is one of the important experimental probes of collective
dynamics in $A+A$ reactions \cite{flow-review}. 
It results from the spatial asymmetry 
in the transverse plane in non-central collisions,
which is largest at early times.
Therefore, elliptic flow is sensitive to the properties of dense matter, 
such as its equation of state 
\cite{Kolb2,Teaney:v2,Huovinen:2001cy,Kolb:2001qz} 
or the effective scattering cross section of partons produced in 
the collisions \cite{Zhang:1999rs,Molnar:v2,Zabrodin:2001rz,Lin:2001zk}. 
Measurements of elliptic flow at high transverse momentum \cite{star} 
provide important constraints
about the density and effective energy loss of partons 
\cite{pQCDv2,Molnar:v2}. 

Charm quark elliptic flow holds the key to resolve a recent controversy
regarding charm dynamics. 
It was pointed out in Ref.~\cite{Batsouli:2002qf} that
both i) initial perturbative QCD charm production without 
final state interactions (as modeled in PYTHIA\cite{Sjostrand:1993yb})
and ii) complete thermal equilibrium for charm hadrons 
are consistent with the single electron spectra from 
PHENIX \cite{Adcox:2002cg}. 
Information on the magnitude of charm quark elliptic flow is essential
to distinguish between these two extreme scenarios.

In this study we determine the elliptic flow of charm mesons and baryons 
based on quark coalescence
and investigate how to extract the charm quark elliptic flow from
that of charm hadrons.
We extend the simple quark coalescence formalism 
to the case of different constituent quark masses,
which is relevant for heavy flavors,
and also consider the effect of the finite momentum width of 
hadron wavefunctions. 

We find that relations between hadron and parton elliptic flow are modified 
for hadrons with constituent quarks of unequal masses. 
As a result, charm hadron elliptic flow is predicted to increase 
much slower with transverse momentum and saturate at a much higher 
momentum scale than the flow of hadrons with only lighter flavors. 
A very slow increase of charm hadron elliptic flow with $\pt$
is a strong indication for a small charm quark elliptic flow.
The effect of the finite momentum spread in hadron wavefunctions 
is relatively small for charm hadrons. 
In this case, simple relations exist to unfold both light and charm quark 
$v_2(\pt)$ from hadron elliptic flow. 
Comparing these predictions with upcoming experimental data 
will tell to what degree charm quarks rescatter in the medium 
and provide valuable insights into the dynamics and 
hadronization of the dense partonic matter in heavy ion collisions.

\section{Mass and wavefunction effects in quark coalescence}
A convenient starting point
to describe meson production via coalescence  $\alpha \beta \to M$
is the relation \cite{Dover:1991zn}
\ber
E \frac{dN(\vp)}{d^3p}\! = \! g_M \!\!
\int\! \! \frac{d\sigma^\mu p_\mu}{(2\pi)^3} \!\!
\int\! \! d^3 \! q \! \left|\psi_\vp \, (\vq)\right|^2 \!
f_\alpha(\vpa,x) f_\beta(\vpb,x) \ 
\label{coalq}
\eer
between the phase space distributions of constituent quarks $\alpha$ 
and $\beta$, the meson wavefunction $\psi_\vp$ \cite{note},
and the invariant momentum distribution of produced mesons.
Here $\vp\equiv\vec {p_\alpha}+\vec {p_\beta}$, 
$\vq\equiv\vec {p_\alpha}-\vec {p_\beta}$, 
$g_M$ is the statistical factor for the meson formation 
\cite{Dover:1991zn,Fries:2003vb,Greco:2003xt}, and the integration runs over a 
3D space-time hypersurface parameterized by $\sigma^\mu(x)$.
The expression is valid if coalescence is a rare process.
When the coalescence probabilities are high for the constituents, 
instead of the quadratic (cubic) scaling of the meson (baryon) number 
with the quark number implied by the above equation, 
the scaling should be linear \cite{Voloshin:2002wa,Molnar:2003ff}. 
Eq.~(\ref{coalq}) also neglects the hadronic binding energy.
Because of the large charm quark mass,
this assumption is better met for charm hadrons than for lighter ones.
For quark coalescence into baryons $\alpha \beta \gamma \to B$,
Eq.~(\ref{coalq}) can be generalized in a straightforward manner.

Note, Eq.~(\ref{coalq}) is valid for collectively expanding sources,
provided the parton distribution functions, $f_i(\vp,x)$, 
change little over the spatial size of the hadron wavefunctions 
($\sim 1$ fm).
However, it  has to be modified 
if flow velocity gradients are very large as shown by Eqs.~(3.19-20) in 
Ref.~\cite{Scheibl:1998tk}.
We leave the discussion 
of this and other space-momentum correlation effects to further studies. 

For a hadron at rest, $\psi_{\vec 0}$ has a small momentum space extension of 
$\lqcd \sim 1/$fm (based on the uncertainty principle)
and therefore Eq.~(\ref{coalq}) reduces to the simple formula
considered in Ref.~\cite{Molnar:2003ff}. 
For a fast moving hadron, on the other hand, 
the wavefunction can change significantly due to Lorentz boost.
We estimate this effect assuming that $\psi_\vp$ 
is dominated by the contributions of dressed valence quarks. 
In this case, e.g., for mesons, $|\psi_{\vec 0}(\vqrest)|^2$ is
the probability density for finding quark/antiquark $\alpha$ and $\beta$ 
with momenta $(\vqa, \vqb)=(+\vqrest/2, -\vqrest/2)$ in the hadron rest frame
(primed quantities refer to the hadron rest frame throughout this study). 
For a hadron at mid-rapidity with momentum $\vp=(\pt \vnt, 0)$ 
in the LAB frame, the transverse momenta of valence quarks 
along the transverse boost direction $\vnt$ are given by
\ber
\pti = \frac{E^\prime_i}{m_M} \pt + \vqi \! \cdot \! \vnt
\frac {\sqrt{\pt^2 + m_M^2}}{m_M} \ .
\label{boost}
\eer
Here $E^\prime_i \equiv (m_i^2 + |\vqi|^2)^{1/2}$, 
$m_i$ is the effective mass of valence quark $i$, and 
$m_M$ is the meson mass.

Suppose now that the rest-frame wavefunction was sufficiently narrow 
so that one can take $\vqrest \rightarrow 0$.
For the weakly bound system assumed, $m_M \approx m_\alpha + m_\beta$ and
Eq.~(\ref{boost}) then gives $\pti = \pt m_i/(m_\alpha+m_\beta)$.
This relation also holds in general for the average constituent momenta,
provided $(\vqrest_i)^2 \ll m_i^2$.
Introducing the constituent momentum fractions
$z_i \equiv \pti/\pt$,
for mesons and baryons we then have, respectively,
\ber
\mzi = \frac{m_i}{m_\alpha+m_\beta} , ~
\mzi = \frac{m_i}{m_\alpha+m_\beta+m_\gamma} .
\label{zi}
\eer
If the effective masses of constituent quarks are similar,
$\bar \pta = \bar \ptb = \pt/2$ for a meson, 
which is the case, e.g., for pions or the $J/\psi$.
On the other hand, for $D$ mesons
$m_\alpha \ll m_\beta$ and therefore 
most of the hadron momentum is carried by the heavy quark.
The asymmetric momentum configuration arises because coalescence requires
the constituents to have similar {\em velocities}, not momenta.

The Lorentz boost also affects the width of the hadron wavefunction. 
From Eq.~(\ref{boost}) the spreads of valence quark momentum fractions
in the LAB frame are
\ber
\delta z_i= \frac{\vqi \! \cdot \! \vnt}{m_H} \frac {\sqrt{\pt^2+m_H^2}}{\pt}
\approx \frac{\vqi \! \cdot \! \vnt}{m_H}
\label{dz}
\eer
if the hadron is moving relativistically.
For massive hadrons, such as charm hadrons,
$\delta z_i \sim \vqi \cdot \vnt/m_H \ll 1$ 
because $\vqi$ is typically on the order of $\lqcd$. 
On the other hand, for hadrons with masses comparable to $\lqcd$, 
e.g., pions, $\delta z_i$ is always 
on the order of unity regardless of the transverse momentum. 
However, note that for such light mesons 
the binding energy cannot be neglected
and hence the conventional coalescence formalism may be not be reliable.

Since the constituent quark momentum components perpendicular to the 
hadron momentum (i.e., $\vnt$ at mid-rapidity) are small ($\sim \lqcd$),  
we further simplify Eq.~(\ref{coalq}) 
by considering only the quark momentum component along $\vnt$.
Then the integrals over the wavefunction can be recast, 
for mesons for example, as  $\int d^3q |\psi_\vp(\vq)|^2 = 
\int dz_\alpha |\Phi_\vp (z_\alpha)|^2$.
As seen from Eq.~(\ref{dz}), $\delta z_i$ 
is independent of $\pt$ for large $\pt$ and thus 
$\Phi_\vp (z_\alpha)$ only depends on $z_\alpha$.
In this case Eq.~(\ref{coalq}) becomes, with $z_\beta=1-z_\alpha$, 
\ber
E \! \frac{dN\! (\vp)}{d^3p}\!\! = \!\! g_M \!\!\!
\int\!\! \frac{d\sigma^\mu p_\mu}{(2\pi)^3} \!\!
\int\!\!\! dz_\alpha \! \left|\Phi_M \! (\! z_\alpha \!)\right|^2 
\!\! f_\alpha (\! z_\alpha \vp,\! x) f_\beta (\!z_\beta \vp,\! x) .
\label{coalz}
\eer

\section{Elliptic flow for charm hadrons}
\subsection{Narrow wavefunction case (the limit of zero momentum spread)}
Because for charm hadrons 
the spread $\delta z_i$ is small,
it is a good approximation to consider only the mean quark momentum fraction 
$\mza$ in Eq.~(\ref{coalz}).
For simplicity, we shall neglect the spatial variation of $f_i$ on the 
hypersurface and assume that in non-central heavy ion collisions
the $\cos (2\phi)$ component is the only non-trivial 
term in the quark azimuthal distribution, i.e.,
\ber
f_i(\vpt) \equiv (2\pi)^3 \frac {dN_i}{d^3p}
 = h_i(\pt) \left [ 1+2 v_{2,i}(\pt) \cos  (2\phi) \right ] .
\label{v2only}
\eer
Eq.~(\ref{zi}) then relates $v_2(\pt)$ of mesons or baryons to 
those of the constituent quarks as \cite{Molnar:2003ff}
\ber
&&v_2^M(\pt)= \frac {\mva+\mvb}{1+2~\mva~ \mvb} ~\simeq \mva+\mvb, \nonumber \\
&&v_2^B(\pt)= \frac {\mva+\mvb+\mvd+3~\mva~ \mvb~ \mvd}
{1+2 ~\left (\mva ~\mvb + \mva~ \mvd + \mvb~ \mvd \right )} \nonumber \\
&&\hspace {1.2cm} \simeq \mva+\mvb+\mvd ,
\label{v2h}
\eer
where $\overline{v_{2,i}} \equiv v_{2,i}(\mzi \pt)$ is the elliptic 
flow of valence quark $i$ at its average momentum.
As a result, for hadrons consisting of both light and charm quarks, 
elliptic flow at a given $\pt$  contains the light quark elliptic flow at 
a much smaller $\pt$. 

If the light quark density is so high that 
the coalescence probability for a charm quark is unity, 
Eq.~(\ref{v2h}) breaks down, and charm hadrons 
just inherit the flow contribution from charm quarks \cite{Voloshin:2002wa}.
Preliminary experimental data indicate that 
$\pt$ scale for light quarks below which this occurs might be as low as 1 GeV 
\cite{star}.
Also note that our study does not include the independent fragmentation 
of partons, which eventually dominates over coalescence 
at very large $\pt$ \cite{Fries:2003vb,Greco:2003xt,Molnar:2003ff}, 
due to the power-law shape of the parton spectra at high $\pt$. 
Since the fragmentation function of heavy quarks in vacuum 
is much harder than that of light quarks, 
the $\pt$ scale above which independent fragmentation starts to dominate  
can be different for charm quarks than those for light quarks. 

\begin{figure}[htpb]
\centerline{\epsfig{file=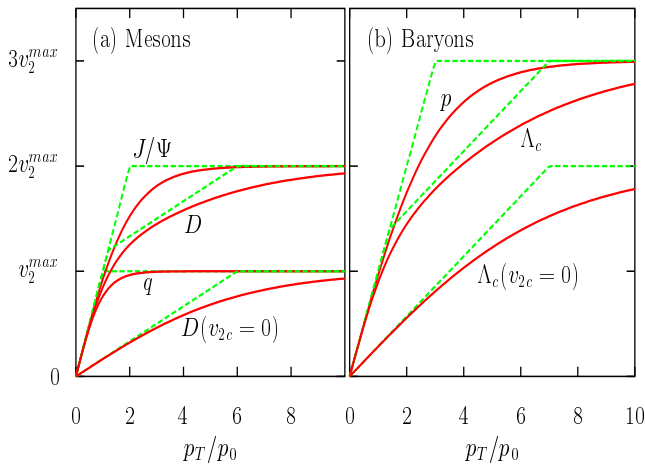,width=3.4in,height=2.5in,angle=0}}
\caption{Comparison of $v_2(\pt)$ for quarks and charm hadrons.}
\label{figv2c}
\end{figure}

Fig.~\ref{figv2c} shows the mass effects on elliptic flow of charm hadrons, 
considering either $v_{2,c}(\pt)=v_{2,q}(\pt)$ (unlabeled) 
or $v_{2,c}(\pt)=0$. 
For illustration, dashed curves correspond to 
a light quark flow $v_{2,q}(\pt)$ which increases linearly then saturates
when $\pt>p_0$. 
Solid lines instead represent a more realistic scenario, where 
$v_{2,q}(\pt)=v^{max}_2 \tanh (\pt/p_0)$,
as found in the Molnar's parton transport model (MPC) \cite{Molnar:v2}.
For $J/\psi$ and protons, which consist of quarks with similar masses, 
$v_2(\pt)$ saturates at $\pt \sim n_v p_0$ at a value 
$n_v v^{max}_2$ \cite{Molnar:2003ff}, 
where $n_v=2$ and $3$ for mesons and baryons, respectively. 
However, for $D$ mesons and $\Lambda_c$ baryons
elliptic flow increases with $\pt$ very slowly  
due to the large difference in the constituent quark masses. 
The reason is that light quarks in these hadrons carry only 
a small fraction of the hadron momentum.
The slow increase is most pronounced when charm quarks have no elliptic flow. 
Furthermore, when charm quarks have elliptic flow, 
$v_2(\pt)$ of $D$ mesons and $\Lambda_c$ baryons 
both exhibit two different slopes because the flow of the heavier 
constituents enters the saturation regime much earlier.
In both cases, $v_2(\pt)$  of these hadrons 
saturates at a scale well above $n_v p_0$. 
These qualitative features could be tested by experiments at RHIC and the LHC,
and they will provide key information on 
the dynamics of charm quarks/hadrons in dense matter created 
in heavy ion collisions \cite{closs,cnoloss,Batsouli:2002qf}. 

\subsection
{Unfolding quark elliptic flow}
When Eq.~(\ref{coalq}) is valid and the effect of the momentum spread is small,
it is possible to unfold the quark elliptic flow from hadron elliptic flow 
using Eq.~(\ref{v2h}). 
For example, if $v_2(\pt)$ is known for $D$ and $\Lambda_c$, 
we have
\ber
&&v^q_2(\pt)=v^{\Lambda_c}_2\left ( (2+r)\pt \right )
-v^D_2 \left ((1+r)\pt \right ), \nonumber \\
&&v^c_2(\pt)=2 v^D_2 \left ( \frac {1+r}{r} \pt \right )
-v^{\Lambda_c}_2 \left ( \frac {2+r}{r} \pt \right ) ,
\eer
where $r \equiv m_c/m_q$ is the ratio of the charm quark and light 
($u$ and $d$) quark effective mass.

\subsection{Numerical estimates including the momentum spreads of wavefunctions}
For rough numerical estimates, 
we consider the conditions in semi-peripheral heavy ion collisions 
at the top RHIC energy,  $\sqrt s=200A$ GeV. 
With the assumption of Eq.~(\ref{v2only}), one obtains
\ber
E \frac{dN(\vp)}{d^3p} &\propto & \int d z_\alpha 
\left|\Phi_M(z_\alpha) \right|^2 
h_\alpha(z_\alpha \pt) h_\beta(z_\beta \pt) \nonumber \\
&& \hspace{-1.9cm} \times \!
\left [ 1\!+\!2 v_{2,\alpha}(z_\alpha \pt \!) \! \cos  (2\phi)  \right ]
\left [ 1\!+\!2 v_{2,\beta}(z_\beta \pt \!) \! \cos  (2\phi)  \right ] . 
~~~~
\eer
We recover Eq.~(\ref{v2h}) when the momentum spread is zero, 
i.e., $\left|\Phi_M(z_\alpha)\right|^2=\delta(z_\alpha-\mza)$. 
Including the momentum spread but neglecting small corrections 
of higher orders in $v_{2,i}$, the meson elliptic flow is 
\ber
v_2^M \! (\pt)\!=\!  
\frac {\int \! d z_\alpha w_M(\! z_\alpha,\pt \! ) 
\left [ v_{2,\alpha}(z_\alpha \pt ) \! + 
\! v_{2,\beta}(z_\beta \pt ) \right ] }
{\int d z_\alpha w_M(z_\alpha,\pt)}, ~~
\label{v2dz}
\eer
where the {\em weight function} for mesons is given by
\ber
w_M(z_\alpha,\pt) =
\left|\Phi_M (z_\alpha) \right|^2 h_\alpha(z_\alpha \pt) h_\beta(z_\beta \pt).
\label{eqwm}
\eer
For baryons, the similar expression for $v_2^B\! (\pt)$ involves 
two integrals, over $z_\alpha$ and $z_\beta$, 
where $w_B(z_\alpha,z_\beta,\pt)$ 
contains the product of three quark distributions. 

For the hadron wavefunctions, for simplicity we take the form 
used in the valon model \cite{Hwa2}:
\ber
\left |\Phi_M(z_\alpha) \right|^2 \propto z_\alpha^a z_\beta^b, ~
\left |\Phi_B(z_\alpha,z_\beta) \right|^2 \propto z_\alpha^a z_\beta^b 
z_\gamma^d,
\label{wavefn}
\eer
where $z_\beta=1-z_\alpha$ for mesons and 
$z_\gamma=1-z_\alpha-z_\beta$ for baryons. 
The normalization constants play no role in Eq.~(\ref{v2dz}) 
for the elliptic flow and have been omitted in the above. 
We take $\lqcd=0.2$ GeV, and constituent quark masses
$m_u=m_d=0.3$ GeV, $m_s=0.5$ GeV, $m_c=1.5$ GeV. 
Observing Eq.~(\ref{dz}) in the large $\pt$ limit, 
we take $\sum {\delta z_i}^2=n_v*(\lqcd/m_M)^2/3$. 
With $\mzi$ given by Eq.~(\ref{zi}), 
the exponents $a$ and $b$ are then determined.
With the convention that $a$ is for the lightest 
and $b$ (or $d$ for a baryon) is for the heaviest quark in a hadron, 
the values of $(a,b)$ are
$(0.25, 1.1)$ for kaons, $(8.2, 8.2)$ for $\phi$, 
$(4.9, 28)$ for $D$, and $(88, 88)$ for $J/\psi$. 
In general, $a$ and $b$ increase for heavier mesons,  
reflecting their narrower spread in $z$. 
For pions the above method gives a singular solution, i.e., 
$a=b<-1$ leading to divergence in 
$\int dz_\alpha \left |\Phi_M(z_\alpha) \right|^2$.
Therefore, for pions we take $a=b=0$ \cite{Hwa2}, 
i.e., a flat distribution in $z$, for simplicity. 
For baryons, the $(a,b,d)$ values are
$(3.6, 3.6, 3.6)$ for protons, $(4.2, 4.2, 7.7)$ for $\Lambda$, 
$(5.3, 9.5, 9.5)$ for $\Xi$, $(14, 14, 14)$ for $\Omega$, 
and $(7.2, 7.2, 40)$ for $\Lambda_c$. 

\begin{figure}[htpb]
\centerline{\epsfig{file=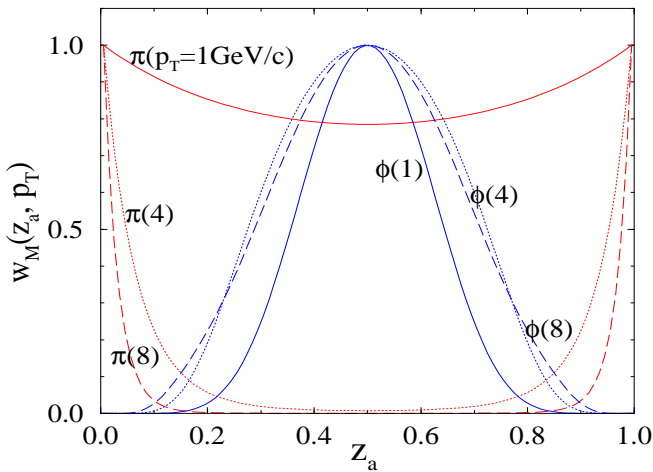,width=3.4in,height=2.5in,angle=0}}
\caption{Weight function $w(\! z_\alpha,\pt \!)$ 
of $\pi$ and $\phi$ at different $\pt$.}
\label{figwm}
\end{figure}

For heavy ion collisions at RHIC energies, the parton momentum distributions 
at mid-rapidity, $h_i(\pt)$, may be parameterized by the sum of an exponential 
soft part and a power-law hard component. 
Assuming these have identical shapes for 
light quarks ($u,d,\bar u,\bar d,s$ and $\bar s$), 
we take $h_i(\pt) \propto e^{-\pt/T} +c_H/(1+\pt/\Lambda_H)^w$ 
with $T=0.18$ GeV, 
$c_H=0.36$, $\Lambda_H=1.3$ GeV, and $w=8.3$. 
For charm (and anti-charm) quarks, we use 
$h_c(\pt) \propto (\pt+0.5$\rm GeV$)^2/(1+\pt/6.8$\rm GeV$)^{21}$,
which reasonably parameterizes the transverse momentum spectrum of 
primary charm quarks at mid-rapidity in $pp$ collisions at $\sqrt s=200$ GeV 
from PYTHIA \cite{Sjostrand:1993yb}. 
We note that medium effects for charm quarks 
in $A+A$ collisions such as energy loss \cite{closs,cnoloss} 
are not considered in this study.

Fig.~\ref{figwm} shows the weight functions (\ref{eqwm}) for pions and 
$\phi$ mesons at $\pt=1$, $4$ and $8$ GeV,
with all maxima normalized to one.  
For the pion, they are not centered around 
the mean value $z_\alpha=0.5$ at higher $\pt$ values.
Instead the dominant momentum configuration for the two constituent quarks 
is around $z_\alpha \sim 0$ and $1$.
This asymmetric momentum configuration at high $\pt$
for wavefunctions that are flat in $z$ is a result of the power-law 
enhancement of the parton spectra at high $\pt$, 
as shown in Ref.~\cite{Lin:2002rw}.
For massive hadrons, such as the $\phi$, 
the wavefunction becomes narrower. 
Actually, for large $a$ and $b$ the meson wavefunction 
in Eq.~(\ref{wavefn}) becomes a Gaussian, 
$\exp [-(z_\alpha-\mza)^2/(2 \delta z_\alpha^2)]$, where 
$\mza \simeq a/(a+b)$ and $\delta z_\alpha^2 \simeq ab/(a+b)^3$.
The narrowness of the wavefunction then dominates over the 
momentum distributions in 
Eq.~(\ref{eqwm}) so that symmetric momentum configurations are usually favored.
Indeed, as shown in Fig.~\ref{figwm}, the weight functions for $\phi$ 
center at $z_\alpha=0.5$ for all three $\pt$ values. 

\begin{figure}[htpb]
\centerline{\epsfig{file=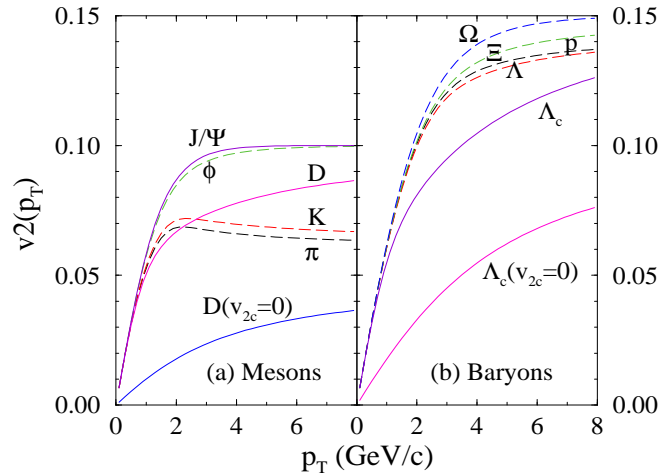,width=3.4in,height=2.5in,angle=0}}
\caption{Elliptic flows of hadrons including momentum spread in wavefunctions: 
a) for mesons and b) for baryons.}
\label{figv2dz}
\end{figure}

We now study the effect of momentum spread on hadron elliptic flow
via Eq.~(\ref{v2dz}) for mesons and the analogous equation for baryons. 
For light quarks including $s$ and $\bar s$, 
the elliptic flow is parameterized as 
$v_{2,q}(\pt)=v^{max}_2 \tanh (\pt/p_0)$ 
with $v^{max}_2=0.05$ and $p_0=0.75$ GeV,
based on parton transport simulations using MPC. 
We consider $v_{2,c}(\pt)=v_{2,q}(\pt)$ unless specified otherwise. 
As shown in Fig.~\ref{figv2dz}, 
all qualitative features for charm hadrons discussed earlier remain valid. 
However, in general hadron elliptic flow is reduced 
relative to Eq.~(\ref{v2h})  
because the concave shape of the quark $v_2(\pt)$ ansatz
tends to penalize any momentum spread
(the average of $v_2(\pt)$ at two different $\pt$ values
is lower than the value at the average $\pt$). 
For example, the value of elliptic flow at $\pt=6$ GeV 
is lower than Eq.~(\ref{v2h}) by 
36\% for pions, 32\% for kaons, 11\%$(24$\%$)$ for $D$ mesons 
with(without) 
charm flow, 9\% for protons, 10\% for $\Lambda$, 5\% for $\Xi$, 
10\%$(17$\%$)$ for $\Lambda_c$ with(without) charm flow. 
These corrections are especially large for pions and kaons at high $\pt$  
and cause the decrease in their $v_2(\pt)$ above $\sim 2$ GeV seen in 
Fig.~\ref{figv2dz}. 
The reason for this suppression is  
that the elliptic flow 
of the slower constituent quark of a pion or a kaon is still far below the 
saturation value
because of the very asymmetric 
momentum configuration \cite{Lin:2002rw} (see Fig.~\ref{figwm}).
However, note that the flows of pions and kaons may be significantly modified 
by resonance contributions and the binding energy, 
which have been neglected in this study.
We also found that the momentum spread has a negligible effect 
on the elliptic flow of massive hadrons, 
especially those with quarks of similar masses. 
The flow reductions are 
0.8\% for $\phi$, 0.01\% for $J/\psi$ and 1\% for $\Omega$, 
thus these hadrons reflect more directly the partonic elliptic flow.

\begin{acknowledgments}
Helpful discussions with U. Heinz, 
B. M\"uller, S. Panitkin and S. Voloshin are greatly appreciated. 
This work was supported by the U.S. Department of Energy under
Grant No. DE-FG02-01ER41190. 
\end{acknowledgments}

\end{document}